\documentclass[useAMS,usenatbib,usegraphicx]{mn2e}
\usepackage{graphicx}
\usepackage{natbib}
\bibpunct{(}{)}{;}{a}{}{,}
\usepackage{journals_aas}   % only needed locally to make .bbl, remove before submission

\def\fracd#1#2{{\displaystyle\frac{#1}{#2}}}

\begin{document}

\title[Robust estimation of orientation parameters]
  {Towards a robust estimation of orientation parameters between ICRF and $Gaia$ celestial reference frames}
\author[Z. Malkin]{Zinovy Malkin$^{1,2}$\\
  $^1$Pulkovo Observatory, St.~Petersburg, 196140, Russia\\
  $^2$Kazan Federal University, Kazan, 420000, Russia}
\maketitle

\begin{abstract}
An analysis of the source position differences between VLBI-based ICRF and $Gaia$-CRF catalogues is a key step
in assessing their systematic errors and determining their mutual orientation.
One of the main factors that limits the accuracy of determination of the orientation parameters between
two frames is the impact of outliers.
To mitigate this effect, a new method is proposed based on pixelization data over the equal-area cells,
followed by median filtering of the data in each cell.
After this, a new data set is formed, consisting of data points near-uniformly distributed over the sphere. 
The vector spherical harmonics (VSH) decomposition is then applied to this data to finally compute
the orientation parameters between ICRF and $Gaia$ frames.
To validate the proposed approach, a comparison was made of the ICRF3-SX and $Gaia$~DR2 catalogues using
several methods for outliers removal.
The results of this work showed that the proposed method is practically insensitive to outliers and thus
provides much more robust results of catalogues comparison than the methods used so far.
This conclusion was confirmed by analogous test comparison of the $Gaia$~DR2 and OCARS catalogues.
\end{abstract}


\section{Introduction}

Constructing and maintenance of the celestial reference system and celestial reference frame (CRF)
is a fundamental task of astrometry.
CRF is realized by catalogues of position of celestial objects in different frequency bands, such as optics,
radio, infrared, and others.
The latest, most accurate CRF realizations in radio is ICRF3 \citep{Charlot2020} obtained from very long baseline
interferometry (VLBI).
The $Gaia$ mission provides an optical CRF of a similar accuracy.
The latest $Gaia$ astrometric solution is $Gaia$ EDR3 \citep{Lindegren2021}, however, the previous release,
$Gaia~DR2$ \citep{Mignard2018}, was used in this study because of its detailed comparison with ICRF3 already
presented in the literature and will be discussed below.

Comparison of astrometric catalogues is a routine procedure used for improvement of the CRF in sense of
stochastic and systematics errors, as well for CRF accuracy assessment.
A detailed overview of the catalogue comparison techniques can be found, e.g., in \citet{Walter2000,Vityazev2017}.
A generally accepted method for catalogue comparison is decomposition of differences between the common object
positions in two catalogues into analytical functions.
Currently, the most widely used method is the vector spherical harmonics (VSH) decomposition
\citep{Makarov2007,Mignard2012,Vityazev2014,Liu2018a}.
Generally speaking, the VSH decomposition is mostly used for simultaneous processing of the catalogues of source
positions and proper motions with allows us to estimate both orientation parameters between frames at the reference
epoch and the speed of time evolution of the orientation parameters.
This technique is naturally applied to star position catalogues containing the source coordinates at the reference
epoch and their linear proper motions.
However, for comparison of the ICRF and $Gaia$-CRF catalogues, only the source positions have been used up to now
because the extragalactic source motions are generally irregular (non-linear) and poorly known, which is mostly
explained by their complicated and variable structure.
For this case of catalogue comparison, which uses only the source position differences, a method can be proposed
to improve the robustness of the estimation of orientation parameters.
Such a method is discussed in this paper.
Generally speaking, this method most probably can be adapted for including the source velocities in the comparison,
but this modification requires a supplement independent study, which is beyond the scope of this paper. 

One of the main problems that can compromise the results of catalogues comparison are outliers.
Outliers are usually present in all measured data, including astrometric positions.
These outliers affect the results of data processing \citep{Titov2013,Frouard2018,Mignard2018,Mayer2020}.
Of course, the authors always try to eliminate outliers before performing the final analysis.
There are numerous approaches to solving this problem.
However, the use of different methods of eliminating outliers leads to different data sets, which, in turn,
gives different results of the analytical representation of the differences between the catalogues.
The impact of outliers becomes more significant with increasing the order of the VSH (or other) decomposition,
and high-order coefficients of the analytical expansion thus became un-reliable.
Therefore, mitigating the impact of outlier on the results of analytical representation of the position
differences between catalogues is an actual astrometric task.

Many different methods for removing outliers were discussed in the literature.
For instance, let us consider several recent papers which include comparisons of the VLBI-based catalogues,
e.g., ICRF2 \citep{Fey2015} or ICRF3, and $Gaia$-based astrometric solutions, such as $Gaia-CRF1$ \citep{Mignard2016} or
$Gaia~CRF2$ \citep{Mignard2018}.
The authors of these papers mostly detected outliers using different criteria based on analysis
of either the distance between source positions in two catalogues $D$ or the normalized separation $X$.
The exact definition of these quantities will be given below.

\citet{Mignard2016} compared $Gaia$-CRF1 and ICRF2 and used two criteria: $X > 4.1$ and $D > 10$~mas.
\citet{Frouard2018} compared usno2016a VLBI solution with ICRF2 and $Gaia$-CRF1 and used a criteria 
$X > 3.67$ and some supplement analysis of the position differences.
\citet{Liu2018b} compared $Gaia$~DR2 and ICRF2 catalogues and used two criteria: $X > 3.97$
and an upper limit of 1 mas on the semi-major axis of the error ellipse.
\citet{Karbon2019} compared several radio source position catalogues with $Gaia$~DR2 and used $X > 4.2$
following \citet{Mayer2018}.
\citet{Makarov2019} compared ICRF3 and $Gaia$~CRF2 and used $X > 4$ and $X > 3$ criteria. 
\citet{Liu2020} compared ICRF1, ICRF2 and ICRF3 with $Gaia$~CRF2 using two criteria:
$X > X_0 = \sqrt{2 \ln N}$, where $N$ is the number of sources, and $D > 10$~mas.
A test computation shows that the value of the first threshold is slowly increasing from $X_0 = 3.90$
to $X_0 = 4.29$ for $N = 2000 \,\ldots\, 10000$, which makes the value of $X_0$ close to those used
by other authors when applying to the current VLBI and $Gaia$ CRF realizations.
\citet{Mayer2020} compared Vienna CRF solutions with $Gaia$~DR2 and used an original procedure
for detecting outliers.
The authors removed each source once from the standard solution and calculated the VSH parameters
using all other sources.
Source that significantly changed one of the VSH parameters (it is checked using a 3-sigma threshold)
\citet{Charlot2020} compared ICRF3-SX with ICRF2 and used two criteria:  $X > 5$ and the semi-major axis
of the error ellipse of the source position in either catalog is larger than 5~mas.

The main purposes of this study is to assess the impact of outliers on the results of estimating the orientation
parameters between source position catalogues, and to investigate possible ways to develop a new robust solution
of this problem with respect to outliers.
One of possible methods to mitigate the impact of outliers discussed in the literature is using pixelization
of the sphere, followed by averaging the data in each pixel (cell).
For example, \citep{Vityazev2014,Vityazev2015,Vityazev2015b} considered pixelization of the data using the equidistant
cylindrical projection (ECP) and HEALPix \citep{Gorski2005} methods to handle catalogues containing
a huge number of stars, and to get a uniform distribution of the data points over the sphere, and mitigate random
errors in the compared catalogues.
In this scheme, original differences in star positions and proper motions were replaced by the cell-averaged values.
However, the numerical experiments performed in the present study have shown that this method does not work well
for VLBI-based source position catalogues because the impact of outliers in the source positions remains strong.
For this reason, the cell median was used in this work instead of the cell average, because the median estimates
is much more robust statistics than the mean.

Another specific feature of this work is using Spherical Rectangular Equal-Area Grid
(SREAG, \citet{Malkin2019,Malkin2020}), which provides a pixelization of a spherical surfcace with equal-area cells
(unlike ECP) near-uniformly distributed over the sphere (more uniformly than HEALPix) with the grid configuration
natural for classical astrometric works.

The proposed strategy was tested with two VLBI-derived catalogues, ICRF3 and
OCARS\footnote{http://www.gaoran.ru/english/as/ac\_vlbi/ocars.txt} \citep{Malkin2018}.
The former catalogue is computed directly from processing of VLBI observations in the framework of activity
of the dedicated IAU Working Group \citep{Charlot2020}, the latter catalogue is a compilation
of published VLBI-based determinations of radio source positions \citep{Malkin2018}.
Both catalogues are compared with $Gaia$~DR2 astrometric solution \citep{Lindegren2018,Mignard2018}.

%%%%%%%%%%%%%%%%%%%%%%%%%%%%%%%%%%%%%%%%%%%%%%%%%%%%%%%%%%%%%%%%%%%%%%%%%%%%%%%%%

\section{Testing different approaches to computation of the orientation parameters between frames}
\label{sect:gaia_icrf3}

Basic orientation parameters between two catalogues (frames) can be defined, in notation of \citet{Mignard2012}, as
\begin{equation}
\begin{array}{rcl}
\Delta\alpha^{\ast} & = & \phantom{-}R_1 \cos\alpha \sin\delta + R_2 \sin\alpha \sin\delta - R_3 \cos\delta \\
                    &   & - G_1\sin\alpha + G_2\cos\alpha \,, \\
\Delta\delta        & = & - R_1 \sin\alpha + R_2 \cos\alpha \\
                    &   & - G_1 \cos\alpha \sin\delta - G_2\sin\alpha\sin\delta + G_3\cos\delta \,. \\
\end{array}
\label{eq:vsh_2}
\end{equation}
where ${\mathbf R} (R_1,R_2,R_3)$ is the rotation vector and ${\mathbf G} (G_1, G_2, G_3)$ is the glide vector.
The six coefficients of this equations are computed by least squares using common sources between the two
compared catalogues.
In this work, the comparison of the $Gaia$~DR2 catalogue with two VLBI-based catalogues ICRF3-SX and OCARS is made.

The ICRF3-SX catalogue contains 4536 sources.
Cross-matching ICRF3-SX and $Gaia$~DR2 sources was made using the search radius of 150~mas, which corresponds
to the value used earlier for cross-matching OCARS and $Gaia$ catalogues \citep{Malkin2018},
and resulted in the list of 3376 common ICRF3-SX/$Gaia$~DR2 sources.
Decreasing the search radius to 100~mas, which was accepted by \cite{Lindegren2018} for matching
ICRF3 prototype with $Gaia$~DR2, resulted in 3373 common ICRF3-SX/$Gaia$~DR2 sources, which gives practically
the same set of common sources as used in this work.

OCARS catalogue version of 11 November 2020 was used in this work.
The catalogue contains 13540 sources, including all 4536 ICRF3-SX sources, 40 ICRF3-XKa sources absent
in the ICRF3-SX catalogue, and 12 ICRF3-K sources absent in both ICRF3-SX and ICRF3-XKa catalogues.
Cross-matching with $Gaia$~DR2 was performed using the search radius of 0.15$''$ for ICRF3 sources,
0.25$''$ for other sources with VLBI-derived positions, and 1$''$ for sources with not reliable positions.
The last group contains only 13 sources and was left in the computations for a variety of data.
Although OCARS being a compiled catalog is less uniform than ICRF, such a comparison is interesting
from the methodological point of view because OCARS has about 1.5 times more common sources
with $Gaia$~DR2, and contains sources with larger position difference with Gaia and more outliers.

The distance between the source positions in two catalogues was computed as
\begin{equation}
D = \sqrt{ {\Delta\alpha^{\ast}}^2 + \Delta\delta^2 }
\label{eq:distance}
\end{equation}
with a simplified estimate of its uncertainty
\begin{equation}
\sigma_D = \sqrt{ \sigma_{\Delta\alpha^\ast}^2 + \sigma_{\Delta\delta}^2 }
\label{eq:sigma_distance}
\end{equation}
where
\begin{eqnarray}
\Delta\alpha^\ast &=& ( \alpha_1 - \alpha_2 ) \cos\delta \,, \nonumber \\[1ex]
\Delta\delta &=& \delta_1 - \delta_2 \,, \nonumber \\[1ex]
\sigma_{\alpha^\ast} &=& \sigma_\alpha \cos\delta \,, \nonumber \\[2ex]
\sigma_{\Delta\alpha^\ast} &=& \sqrt{ \sigma_{\alpha_1^\ast}^2 + \sigma_{\alpha_2^\ast}^2 } \,, \nonumber \\[2ex]
\sigma_{\Delta\delta} &=& \sqrt{ \sigma_{\delta_1}^2 + \sigma_{\delta_2}^2 } \,. \nonumber
\end{eqnarray}

The normalized separation $X$ was computed according to \citet[Eq.~(4)]{Mignard2016}.

For 203 (2.5\%) common OCARS/$Gaia$~DR2 sources which have only coordinates in OCARS without uncertainty
and correlation between right ascension and declination, missing values were computed as uniformly distributed
random numbers in the range 0.01 to 10~ms for right ascension uncertainty, 0.1 to 100~mas for declination
uncertainty, and -0.99 to 0.99 for correlation.
The average level of added uncertainty was chosen somewhat higher than the uncertainty level
in the catalogues used in OCARS because the OCARS sources with incomplete data are usually taken from less
reliable catalogues.
Notice that this procedure was applied to the OCARS data only for the purpose of the present study;
the missing data in the original OCARS catalogue have zero values until they can be updated with actual data.

\begin{figure*}
\includegraphics[clip,width=\textwidth]{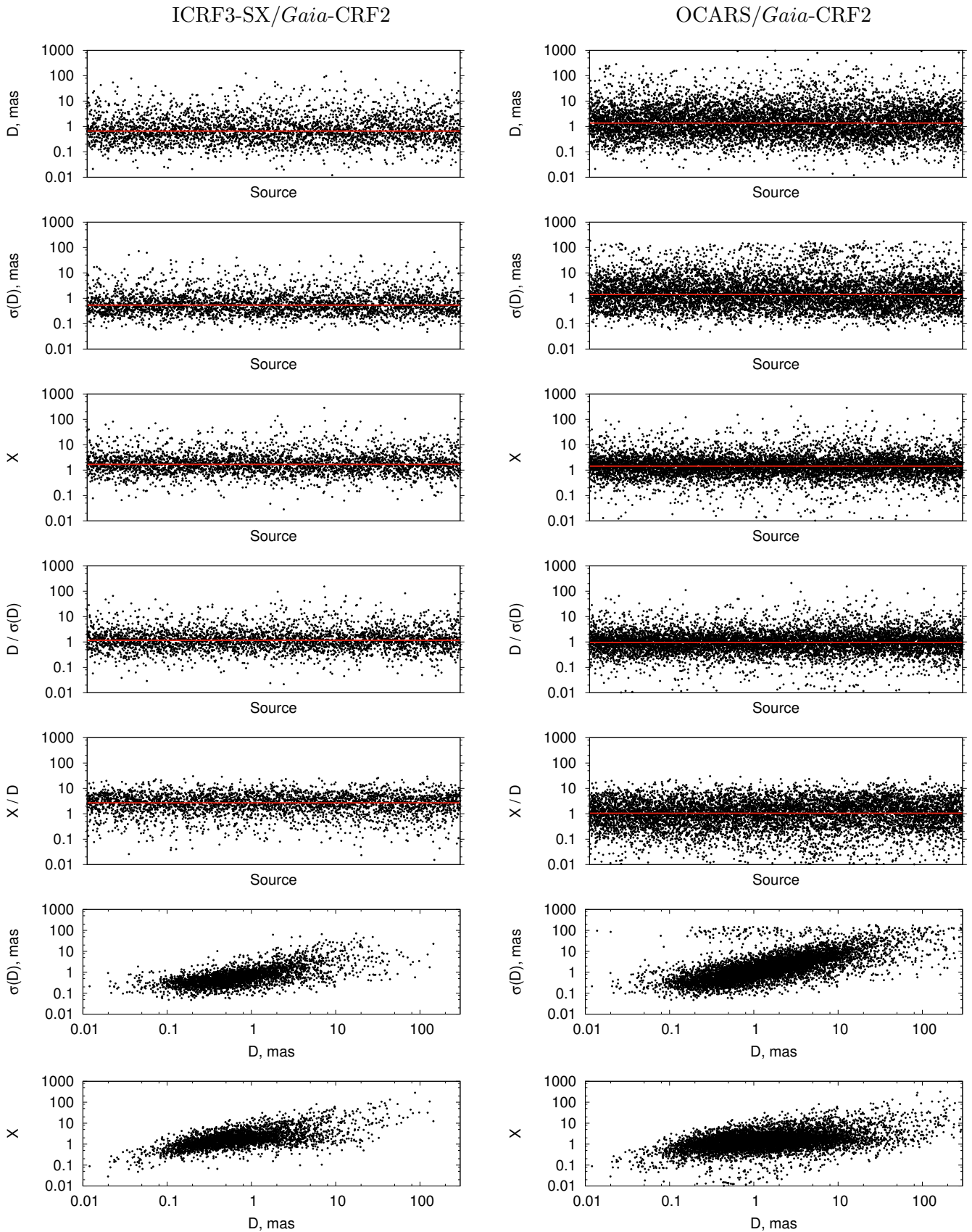}
\caption{Statistics related to the angular separation between source positions in ICRF3-SX and $Gaia$~DR2
  catalogues (left column), and OCARS and $Gaia$~DR2 catalogues (right column): $D$ - distance,
  $\sigma(D)$ - distance uncertainty, X - normalized separation.
  Red lines show the median values.}
\label{fig:dist_sdist_normsep}
\end{figure*}

Figure~\ref{fig:dist_sdist_normsep} shows various statistics related to the position differences
between compared catalogues.

This study is based on the distribution of common sources between the compared catalogues
by equal-area cells on the celestial sphere using the Spherical Rectangular Equal-Area Grid
(SREAG, \citet{Malkin2019,Malkin2020}).
The spatial resolution of this grid is defined by the number of rings $N_{ring}$.
Obviously, the grid resolution (the cell size) should be chosen in such a way that each cell contains
a sufficient number of sources. 
Table\ref{tab:n_ring_max} shows the maximum $N_{ring}$ for which each cell contains at least $N_{min}$ sources.
The data in the Table were computed from the actual source distribution over the sky.

\begin{table}
\centering
\caption{Maximum possible number of rings $N_{ring}$ and cells $N_{cell}$ (which is a function of $N_{ring}$)
  in the grid to ensure the required minimum number of sources $N_{min}$ in each cell
  for ICRF3-SX/$Gaia$~DR2 and OCARS/$Gaia$~DR2 common sources.}
\label{tab:n_ring_max}
\begin{tabular}{ccccc}
\hline
$N_{min}$ & \multicolumn{2}{c}{ICRF3/$Gaia$} & \multicolumn{2}{c}{OCARS/$Gaia$} \\
 & $N_{ring}$ & $N_{cell}$ & $N_{ring}$ & $N_{cell}$ \\
\hline
~1 & 14 & 250 & 18 & 412 \\
~2 & 12 & 184 & 16 & 326 \\
~3 & 10 & 128 & 12 & 184 \\
~4 & 10 & 128 & 12 & 184 \\
~5 & 10 & 128 & 10 & 128 \\
~6 & 10 & 128 & 10 & 128 \\
~7 & ~8 & ~82 & 10 & 128 \\
~8 & ~8 & ~82 & 10 & 128 \\
~9 & ~8 & ~82 & 10 & 128 \\
10 & ~6 & ~46 & 10 & 128 \\
\hline
\end{tabular}
\end{table}

Another important consideration that should be taken into account when choosing $N_{ring}$ is
a desirable maximum order of the VSH expansion $l_{max}$, which is connected with $N_{ring}$ as
$l_{max} = N_{ring}/2$.
Since only the basic orientation parameters corresponding to $l$=1 are computed here, this factor does not
limit the choice of the grid resolution used for this task.

For further computations, the grid with $N_{ring}$=10 ($N_{cell}$=128) was used, which
provides at least six sources in each cell for both comparisons of $Gaia$~DR2 with ICRF-SX and OCARS catalogues.

The actual observed differences between source positions in two compared catalogues may contain
outliers.
Various strategies for detecting and removing outliers are discussed in the literature, as mentioned above
The following criteria, both used in the previous works and combined ones, were tested in this study
to detect outliers (except the first case where outliers are not detected):\\
(1) all the sources are used;\\
(2) distance $D$ is greater than 10~mas;\\
(3) $D$ or its uncertainty $\sigma_D$ is greater than 10~mas;\\
(4) normalized separation $X$ is greater than 4;\\
(5) either of (2) or (4) is met;\\
(6) either of (3) or (4) is met.

The orientation parameters between frames were computed by least squares in two variants.
In the first variant, each common source left after removing outliers was used to form an equation of condition.
In the second variant, median values for each cell in the SREAG grid were used to form the equations of condition;
no preliminary detection and rejection of outliers was performed.
For each of these two variants six criteria for rejecting outliers described above were applied, which gives in total
12 variants of processing. 
All the 12 variants of processing are summarized in Table~\ref{tab:variants}.

\begin{table*}
\begin{center}
\caption{Variants of processing. The first six variants use separate sources, the last six variants use
  the cell median approach. The right part of the table shows the number of sources left after rejecting
  outliers using criteria defined in the left part of the table.}
\label{tab:variants}
\begin{tabular}{ccccccccccc}
\hline
Variant & $D_{max}$ & $\sigma_{D_{max}}$ & $X_{max}$ & $N_{cell}$ & \multicolumn{3}{c}{ICRF3-SX/$Gaia$~DR2} &
\multicolumn{3}{c}{OCARS/$Gaia$~DR2} \\
%\cline{6-8} \cline{9-11}
 & mas & mas & & & $N_{sou}$ & $D_{med}$ & $D_{wrms}$ & $N_{sou}$ & $D_{wrms}$ & $D_{med}$ \\
 & & & & & &mas & mas & & mas & mas \\
\hline
~1 & --- & --- & --- & --- & 3376 (100\%) & 0.664 & 1.809 & 8197 (100\%) & 1.361 & 2.522 \\
~2 & 10  & --- & --- & --- & 3257 (~96\%) & 0.630 & 1.052 & 7446 (~91\%) & 1.163 & 1.111 \\
~3 & 10  & 10  & --- & --- & 3218 (~95\%) & 0.622 & 1.052 & 7060 (~86\%) & 1.111 & 1.110 \\
~4 & --- & --- & ~4  & --- & 2834 (~84\%) & 0.558 & 0.438 & 7287 (~89\%) & 1.206 & 0.611 \\
~5 & 10  & --- & ~4  & --- & 2796 (~83\%) & 0.547 & 0.434 & 6833 (~83\%) & 1.094 & 0.580 \\
~6 & 10  & 10  & ~4  & --- & 2757 (~82\%) & 0.540 & 0.433 & 6447 (~79\%) & 1.033 & 0.578 \\[1mm]
~7 & --- & --- & --- & 128 & & & & & & \\
~8 & 10  & --- & --- & 128 & & & & & & \\
~9 & 10  & 10  & --- & 128 & & & & & & \\
10 & --- & --- & ~4  & 128 & & & & & & \\
11 & 10  & --- & ~4  & 128 & & & & & & \\
12 & 10  & 10  & ~4  & 128 & & & & & & \\
\hline
\end{tabular}
\end{center}
\flushleft{\bf Notes:}\\
1. $D_{max}$ is the tolerance for the distance between source positions in two catalogues,
   $\sigma_{D_{max}}$ is the tolerance for the distance uncertainty,
   $X_{max}$ is the tolerance for the normalized separation,
   $N_{cell}$ is the number of cells in the grid used to compute the cell-averaged differences,
   $N_{sou}$ is the number of common sources in two catalogues used for computation,
   $D_{med}$ is the median distance, and
   $D_{wrms}$ is the WRMS distance.\\
2. Data in columns 6--11 for variants 7--12 are the same as corresponding data for variants 1--6.
\end{table*}

The median is known as a robust statistics much less influenced by outliers than the mean.
Let $\bar{x}_m$ be the median of $x_i$, i.e. $\bar{x}_m = med\{x_i\}$.
A possible approach to compute median uncertainty was proposed by \citet{Mueller2000}.
First, the median of the absolute deviations ($MAD$) is computed as
\begin{equation}
MAD = med\{|x_i - \bar{x}_m|\}.
\label{eq:mad}
\end{equation}
The uncertainty of the median $\bar{x}_m$ is then computed as
\begin{equation}
\sigma_m = \fracd{1.858}{\sqrt{n-1}} \; MAD \,.
\label{eq:sigma_m}
\end{equation}

Notice that this estimate of the median uncertainty depends only on the data scatter and not on the
uncertainties of the data.
As a matter of fact, \citet{Mueller2000} proposed a method of using the uncertainties in the input data
to compute the weighted median and its STD.
However, its practical realization, as pointed out by the author, is more cumbersome, and the testing results
along with the discussion given therein do not show clear advantage of using weighted median.

\begin{figure*}
\centering
\includegraphics[clip,width=\textwidth]{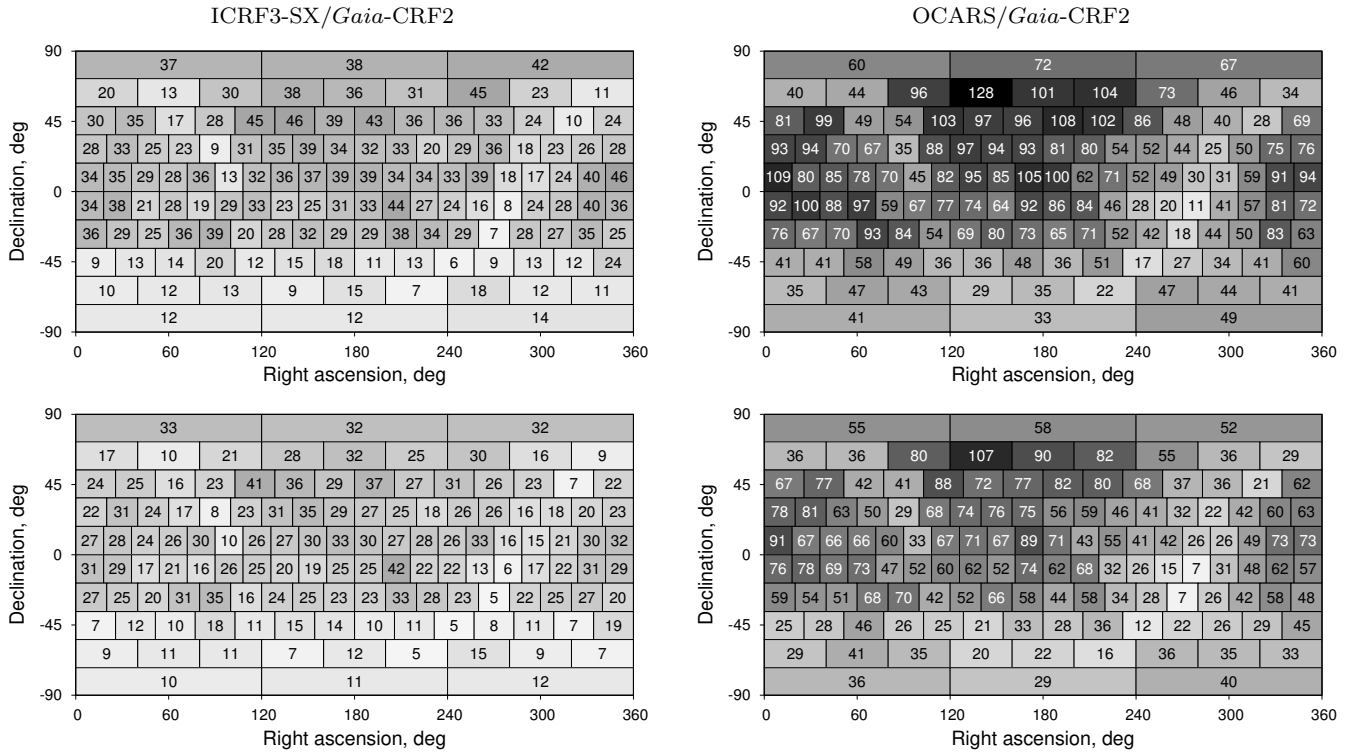}
\caption{Sky distribution of ICRF3-SX/$Gaia$~DR2 (left) and  OCARS/$Gaia$~DR2 (right) common sources:
 top row~-- all sources, bottom row~-- sources left after applying the most thorough rejection criteria,
 see Table~\ref{tab:variants}.
 Cells are labelled with the number of sources fallen into the cell.
 All four plots are made in the same gray scale.}
\label{fig:source_distribution_maps}
\end{figure*}

The common source distribution over the sky is shown in Fig.~\ref{fig:source_distribution_maps}.
Since the $Gaia$ sources are distributed over the sky nearly uniformly, the sky distribution of the common $Gaia$/VLBI
sources mostly reflects the uniformity of the sky distribution of VLBI-based catalogues.
On can see not only general deficit of common sources in the south, but also a deficit of sources in some regions
of the Northern sky too, primarily in the Galactic equatorial zone.

For computation of the orientation parameters between two frames using the cell median method, in fact,
a new data set is used which consists of $N_{cell}$ ``artificial'' common sources with positions equal
to the coordinates of the cells. 
However, the actual distribution of sources in each cell is generally uneven.
Therefore, the coordinates of the geometric center of the cell are not equal to the average coordinates
of the sources that fall into the cell.
A special test was performed to estimate the impact of this effect on the results of computation of the orientation
parameters.
The six components of ${\mathbf R}$ and ${\mathbf G}$ vectors for $Gaia$--ICRF link and $Gaia$--OCARS link
were set to a value of 1~mas and the actual differences between source coordinates in two catalogues
were replaced by computed values.
The coordinates of the sources and, consequently, their distribution over the sky were kept as given in original catalogs.
After that, the six orientation parameters were estimated first using the geometric coordinates of the cells,
and then with the average coordinates of the sources in the cells.
The results of this test presented in Table~\ref{tab:cell_center_coordinates}
show that the orientation parameters are very close.
The difference between two variants is below 1\%, which can be considered as negligible taking into account
that the orientation parameters between the recent CRF realizations is usually within a few tens
of microarcseconds with uncertainty at a level of 10--20\%.
Obviously, the more sources are in the compared catalogues and the more evenly they are distributed over the sky,
the less the impact of choosing the cell center coordinates, and, finally, the more accurate the result is.

\begin{table}
\centering
\caption{Test of the impact of choice of the determination of coordinates of the cell centers. Unit: $\mu$as.}
\label{tab:cell_center_coordinates}
\begin{tabular}{crrrrrrc}
\hline
\multicolumn{1}{c}{$R_1$} & \multicolumn{1}{c}{$R_2$} & \multicolumn{1}{c}{$R_3$} &
\multicolumn{1}{c}{$G_1$} & \multicolumn{1}{c}{$G_2$} & \multicolumn{1}{c}{$G_3$} \\
\hline
\multicolumn{6}{c}{ICRF3/$Gaia$, geometric cell centers} \\
~995 & ~999 & 1000 & 1000 & ~991 & 1001 \\
\multicolumn{6}{c}{ICRF3/$Gaia$, average cell centers} \\
~995 & ~999 & 1000 & ~999 & ~994 & 1000 \\
\multicolumn{6}{c}{OCARS/$Gaia$, geometric cell centers} \\
~997 & ~998 & 1001 & ~999 & ~992 & 1002 \\
\multicolumn{6}{c}{OCARS/$Gaia$, average cell centers} \\
~996 & ~999 & 1000 & ~998 & ~994 & 1000 \\
\hline
\end{tabular}
\end{table}

Full results of the test computations for all the variants are shown in
Table~\ref{tab:orienttaion_angles_icrf} for $Gaia$--ITRF link, and
Table~\ref{tab:orienttaion_angles_ocars} for $Gaia$--OCARS link. 

\begin{table*}
\begin{center}
\caption{Orientation parameters between ICRF3-SX and $Gaia$~DR2 depending on the analysis options. Units: $\mu$as.}
\label{tab:orienttaion_angles_icrf}
\begin{tabular}{cr@{$\pm$}rr@{$\pm$}rr@{$\pm$}rr@{$\pm$}rr@{$\pm$}rr@{$\pm$}r}
\hline
Variant & \multicolumn{2}{c}{$R_1$} & \multicolumn{2}{c}{$R_2$} & \multicolumn{2}{c}{$R_3$}
        & \multicolumn{2}{c}{$G_1$} & \multicolumn{2}{c}{$G_2$} & \multicolumn{2}{c}{$G_3$} \\
\hline
~1a & $-11$ & 30 & $ 27$ & 28 & $ 41$ & 28 \\
~2a & $-12$ & 17 & $ 44$ & 16 & $ 15$ & 16 \\
~3a & $-12$ & 17 & $ 44$ & 16 & $ 15$ & 16 \\
~4a & $-21$ &  7 & $ 24$ &  7 & $ -4$ &  7 \\
~5a & $-21$ &  7 & $ 24$ &  7 & $ -4$ &  7 \\
~6a & $-21$ &  7 & $ 24$ &  7 & $ -4$ &  7 \\
~7a & $-26$ & 13 & $ 45$ & 13 & $-13$ & 12 \\
~8a & $-27$ & 13 & $ 48$ & 13 & $-12$ & 12 \\
~9a & $-26$ & 13 & $ 45$ & 13 & $-14$ & 12 \\
10a & $-26$ & 13 & $ 42$ & 12 & $-15$ & 11 \\
11a & $-24$ & 13 & $ 43$ & 12 & $-15$ & 11 \\
12a & $-25$ & 13 & $ 43$ & 12 & $-14$ & 11 \\
~1b & $-23$ & 31 & $ 19$ & 29 & $ 42$ & 28 & $-29$ & 30 & $ 39$ & 29 & $ 14$ & 30 \\
~2b & $-17$ & 17 & $ 40$ & 17 & $ 16$ & 16 & $-15$ & 17 & $ 16$ & 16 & $  9$ & 17 \\
~3b & $-17$ & 18 & $ 40$ & 17 & $ 16$ & 16 & $-15$ & 17 & $ 16$ & 16 & $  9$ & 17 \\
~4b & $-18$ &  8 & $ 23$ &  7 & $ -4$ &  7 & $ -8$ &  7 & $-10$ &  7 & $ 14$ &  7 \\
~5b & $-17$ &  8 & $ 23$ &  7 & $ -4$ &  7 & $ -8$ &  7 & $-10$ &  7 & $ 14$ &  7 \\
~6b & $-17$ &  8 & $ 23$ &  7 & $ -4$ &  7 & $ -8$ &  7 & $-10$ &  7 & $ 14$ &  7 \\
~7b & $-20$ & 14 & $ 47$ & 13 & $-14$ & 12 & $  6$ & 13 & $-14$ & 12 & $ 19$ & 13 \\
~8b & $-22$ & 14 & $ 50$ & 13 & $-13$ & 12 & $  6$ & 13 & $ -9$ & 12 & $ 23$ & 13 \\
~9b & $-21$ & 14 & $ 47$ & 13 & $-14$ & 12 & $  3$ & 13 & $-10$ & 12 & $ 22$ & 13 \\
10b & $-18$ & 13 & $ 44$ & 13 & $-16$ & 11 & $  5$ & 13 & $-22$ & 12 & $ 27$ & 13 \\
11b & $-17$ & 13 & $ 46$ & 12 & $-16$ & 11 & $  8$ & 12 & $-20$ & 12 & $ 28$ & 12 \\
12b & $-19$ & 13 & $ 45$ & 12 & $-14$ & 11 & $  6$ & 12 & $-18$ & 12 & $ 28$ & 13 \\
\hline
\end{tabular}
\end{center}
\flushleft{\bf Notes:}\\
  1. In variants 1a--12a, only the rotation angles between two frames were estimated;
     in variants 1b--12b, both rotation and glide parameters were estimated. \\
  2. Variants 7--12 based on cell medians are given for the case of the geometric cell center coordinates,
     which are practically the same as the results obtained with the average cell center coordinates.
\end{table*}

\begin{table*}
\centering
\caption{The same as Table~\ref{tab:orienttaion_angles_icrf} for comparison of OCARS and $Gaia$~DR2.}
\label{tab:orienttaion_angles_ocars}
\begin{tabular}{cr@{$\pm$}rr@{$\pm$}rr@{$\pm$}rr@{$\pm$}rr@{$\pm$}rr@{$\pm$}r}
\hline
Variant & \multicolumn{2}{c}{$R_1$} & \multicolumn{2}{c}{$R_2$} & \multicolumn{2}{c}{$R_3$}
        & \multicolumn{2}{c}{$G_1$} & \multicolumn{2}{c}{$G_2$} & \multicolumn{2}{c}{$G_3$} \\
\hline
~1a & $-18$ & 27 & $ 35$ & 25 & $ 38$ & 25 \\
~2a & $-12$ & 12 & $ 46$ & 11 & $ 13$ & 11 \\
~3a & $-12$ & 12 & $ 47$ & 11 & $ 13$ & 11 \\
~4a & $-24$ &  6 & $ 28$ &  6 & $ -5$ &  6 \\
~5a & $-24$ &  6 & $ 28$ &  6 & $ -5$ &  6 \\
~6a & $-24$ &  6 & $ 28$ &  6 & $ -5$ &  6 \\
~7a & $-33$ & 14 & $ 47$ & 14 & $-14$ & 12 \\
~8a & $-31$ & 14 & $ 52$ & 13 & $-14$ & 11 \\
~9a & $-33$ & 14 & $ 51$ & 13 & $ -9$ & 11 \\
10a & $-40$ & 13 & $ 44$ & 12 & $-16$ & 10 \\
11a & $-34$ & 13 & $ 47$ & 12 & $-11$ & 10 \\
12a & $-36$ & 13 & $ 45$ & 12 & $ -7$ & 11 \\
~1b & $-29$ & 28 & $ 22$ & 27 & $ 41$ & 25 & $-46$ & 27 & $ 38$ & 25 & $ 27$ & 27 \\
~2b & $-20$ & 12 & $ 42$ & 12 & $ 14$ & 11 & $-18$ & 12 & $ 25$ & 11 & $ 12$ & 12 \\
~3b & $-20$ & 13 & $ 42$ & 12 & $ 14$ & 11 & $-18$ & 11 & $ 25$ & 11 & $ 12$ & 12 \\
~4b & $-23$ &  7 & $ 25$ &  6 & $ -4$ &  6 & $-13$ &  6 & $  0$ &  6 & $ 17$ &  6 \\
~5b & $-23$ &  7 & $ 25$ &  6 & $ -4$ &  6 & $-12$ &  6 & $  0$ &  6 & $ 17$ &  6 \\
~6b & $-23$ &  7 & $ 25$ &  6 & $ -4$ &  6 & $-12$ &  6 & $  0$ &  6 & $ 16$ &  6 \\
~7b & $-45$ & 15 & $ 45$ & 15 & $-15$ & 11 & $ -9$ & 13 & $ 25$ & 13 & $ 19$ & 15 \\
~8b & $-39$ & 14 & $ 51$ & 13 & $-15$ & 11 & $ -5$ & 13 & $ 20$ & 12 & $  2$ & 14 \\
~9b & $-39$ & 14 & $ 50$ & 14 & $-10$ & 11 & $ -3$ & 13 & $ 16$ & 12 & $  1$ & 14 \\
10b & $-50$ & 14 & $ 44$ & 13 & $-17$ & 10 & $ -3$ & 12 & $ 21$ & 12 & $ 15$ & 13 \\
11b & $-41$ & 14 & $ 45$ & 13 & $-12$ & 10 & $ -5$ & 12 & $ 16$ & 12 & $  4$ & 13 \\
12b & $-40$ & 14 & $ 45$ & 13 & $ -8$ & 11 & $ -1$ & 12 & $ 13$ & 12 & $  2$ & 13 \\
\hline
\end{tabular}
\end{table*}

Analysis of these tables allows us to make the following observations:
\begin{itemize}
\item The rotation angles computed without glide adjustment (variants marked with `a') are close to those
  computed with glide vector (variants marked with `b').
\item Variants with using normalized separation criterion for removing outliers (variants 4--6) systematically differ
  from variants with using $D_{max}$ and $\sigma_{D_{max}}$ criteria in both parameters value and their uncertainty.
\item Variants 4-6 showed practically the same results.
  This means that after applying the normalized separation criterion the distance and distance uncertainty criteria
  (in tested, practically used cases) does not have any significant effect.
\item Variants 2 and 3, as well as variants 5 and 6 give practically identical results.
  This means that applying $\sigma_{D_{max}}$ criterion in addition to $D_{max}$ criterion does not have any
  significant effect.
\item The uncertainties in variant 4 are smaller than those in variants 1--3, evidently because of
  the fewer sources used in the computation, which have a smaller scatter compared with the previous variants.
  Variants 4--6 also give practically identical results because very close set of sources is used in these variants.
\item Variants with using of cell medians show better agreement in both orientation angles and their uncertainties
  between six variants with different approaches to removing outliers, and even without preliminary removing outliers.
\item For comparison the $Gaia$~DR2 catalogue with both ICRF-SX and OCARS, the uncertainty of the orientation
  parameters for variant (1) is about two times larger than those for variants (2) and (3) and about four times large than
  for variants (4)--(6).
  In both cases, variants (7)-(12) computed with the cell median position differences show practically the same
  uncertainties of the orientation parameters for all options for removing outliers.
\item The uncertainties of the orientation parameters obtained using classical source-wise approach differ between
  variants 1--6, i.e. depend on the method of detecting outliers.
  Uncertainties of the orientation parameters obtained using cell-wise method are practically the same for all six
  variants 7--12.
\item The uncertainties of the orientation parameters obtained in variants 7--12 are larger than
  uncertainties for variants 4--6.
  The primary reason of this is the different number of equations of condition in the LS solution, which is the number
  of common sources for the traditional approach and the number of cells for the cell median method.
\end{itemize}

Computation of the orientation parameters from cell medians can provide quite precise result, practically free
of the impact of outliers.
However, as was mentioned above, the uncertainties of the orientation parameters obtained by cell-based method
are substantially larger than the uncertainties obtained by traditional approach based on the outlier detection
with using the $X$ criterion.
This can be considered as a price paid for robustness.
On the other case, the uncertainties obtained by cell-based method may be considered as more realistic.
This effect will decrease with increasing of the number of cells used (for the same number of sources).

%%%%%%%%%%%%%%%%%%%%%%%%%%%%%%%%%%%%%%%%%%%%%%%%%%%%%%%%%%%%%%%%%%%%%%%%%%%%%%%%%%%%%

\section{Conclusion}

In this study, a new method is proposed to improve the robustness of the estimates of the orientation
parameters between two CRF realizations, in particular between the VLBI-based ICRF and $Daia$-CRF.
The method is based on the median averaging of the catalogue position differences over the equal-area sky cells
using SREAG grid \cite{Malkin2019,Malkin2020}.
Several test with real data have proved that the orientation model parameters derived with the new method 
are practically insensitive to outliers.
The results of this study can be applied to any method of analytical representation of systematic differences
in position between two catalogues.

The results presented in this paper were computed with the 128-cell SREAG grid.
Some computations were also made with 82-cell and 184-cell grids.
The results obtained in these tests are similar to those obtained in Section~\ref{sect:gaia_icrf3},
which confirms the conclusions discussed above.

Another advantage of using the proposed technique is mitigation of peculiar apparent source movements caused
by radio and optical source structure effects and their impact on the stability of the parameters of
the orientation between CRF realizations.

From the latter and some other considerations, it is important to choose the optimal number of cells in the grid.
A smaller number of cells ensures more sources fall into each cell, and thus helps to eliminate or at least mitigate
the effects of outliers, source movement and others.
On the other hand, increasing the number of cells allows us to apply a higher order of decomposition into spherical
functions and investigate the finer details of systematic differences between frames.
For SREAG, $N_{ring}$ corresponds to the maximum order of the VSH decomposition, which simplifies the selection
of the optimal grid resolution that is most suitable for given task.
Generally speaking, a higher order of VSH expansion requires a larger number of cells in the celestial grid,
which in turn requires a larger number of sources sufficiently uniformly distributed over the sky.

The proposed technique can also be used to mitigate the impact of outliers in other astrometric works.
For example such an approach can be used for determination of the Galactocentric acceleration
of the solar system from VLBI- or $Gaia$-based source velocity field \citep{MacMillan2005,Titov2013,Klioner2020}.

Other statistical techniques can be also applied to compute a robust estimate of the mean and its realistic
uncertainty as discussed in \citet{Malkin2012}, but it is not expected that the results will be significantly
different.
The median looks the simplest and most effective statistics for solving the problem considered in this work.

Generally speaking, the method discussed in this paper may be also applicable to analysis of astrometric catalogues
including velocity field, but this requires a special investigation.

Grid averaging procedure is analogous to the concept of \textit{normal points} which was widely used in optical astrometry.
It also allows us to discuss a new way to define the system of a celestial frame realization.
Currently, the ICRF orientation is fixed by a set of \textit{defining} sources of high astrometric quality and nearly
uniformly distribute over the sky.
For example, ICRF3 contains 303 defining sources \citep{Charlot2020}.
The main problem with the concept of defining sources is that there are not enough such sources.
Instead, we can consider the predefined conventional celestial grid for which the cell median differences
will define a mutual orientation between frames.
This method will also allow us to mitigate the impact of source structure, including its variability.
With an increase in the number of ICRF sources, the impact of the peculiarities of individual sources
will be more and more mitigated if the cell median approach will be used. 

The results of this work also allow us to confirm and quantify generally known conclusions about the accuracy
and reliability of the ICRF and $Gaia$-CRF comparison.
On can see not only a general deficit of ICRF sources in the south, but also a deficit of optically
bright ICRF sources, which can be used for comparison and link between ICRF and $Gaia$~DR2.
A deficit of common ICRF3/$Gaia$ sources can be also observed in some regions of the Northern sky,
mostly near the Galactic equator, as shown in Fig.~\ref{fig:source_distribution_maps}.
OCARS catalogue \citep{Malkin2018} is designed, in particular, to help in solving this problem.
This catalogue includes all the VLBI-detected radio sources with published accurate coordinates along with its
optical properties, and thus can be used to search for prospective common ICRF/$Gaia$ sources.
Filling in of the regions bare of common ICRF/$Gaia$ sources with new ICRF sources will allow to obtain a more
systematically accurate estimates of high-order harmonics in the differences between optical and radio frames.
This proposal was discussed in more detail in \cite{Malkin2021a}.

\section*{Acknowledgements}

Constructive comments and useful suggestions from an anonymous reviewer are gratefully acknowledged.
The author is grateful to Sergey Klioner for his help in better understanding the $Gaia$ astrometric solution
and using the $Gaia$ archive.

This work has made use of data from the European Space Agency (ESA) mission $Gaia$\footnote{https://www.cosmos.esa.int/gaia},
processed by the $Gaia$ Data Processing and Analysis
Consortium\footnote{https://www.cosmos.esa.int/web/gaia/dpac/consortium} (DPAC).
Funding for the DPAC has been provided by national institutions, in particular the institutions participating
in the $Gaia$ Multilateral Agreement.

This research has made use of the SAO/NASA Astrophysics Data System\footnote{https://ui.adsabs.harvard.edu/} (ADS).
The figures were prepared using \verb"gnuplot"\footnote{http://www.gnuplot.info/}.

\section*{Data availability}

ICRF3 catalogue is available at the website of the International Earth Rotation and Reference Systems Service (IERS),
\textit{https://hpiers.obspm.fr/icrspc/newwww/index.php}.
$Gaia$ catalogue is available at \textit{https://gea.esac.esa.int/}.
OCARS catalog and Fortran routines to perform basic operations with SREAG are available at
\textit{http://www.gaoran.ru/english/as/ac\_vlbi/} and \textit{https://github.com/zmalkin4gt/SREAG}.

\bibliography{zm_orientation}
\bibliographystyle{mn}

\end{document}